\documentclass[twocolumn,aps,showpacs,prl,superscriptaddress]{revtex4-1}
\usepackage[T1]{fontenc}
\usepackage[utf8]{inputenc}
\usepackage{amsmath}
\usepackage{amssymb}
\usepackage{esint}
\usepackage{graphicx}

\usepackage{color}
\usepackage{subfigure}
\usepackage[normalem]{ulem}

\begin{document}
\title{Long-lived non-thermal states realized by atom losses in one-dimensional quasi-condensates}
\author{A. Johnson}
\affiliation{Laboratoire Charles Fabry, Institut d'Optique, CNRS, Universit\'e Paris
Sud 11, 2 Avenue Augustin Fresnel, F-91127 Palaiseau Cedex, France}
\author{S. S. Szigeti}
\affiliation{School of Mathematics and Physics,  University of Queensland, Brisbane, QLD, 4072, Australia}
\affiliation{ARC Centre of Excellence for Engineered Quantum Systems, University of Queensland, Brisbane, QLD 4072, Australia}
\author{M. Schemmer}
\affiliation{Laboratoire Charles Fabry, Institut d'Optique, CNRS, Universit\'e Paris
Sud 11, 2 Avenue Augustin Fresnel, F-91127 Palaiseau Cedex, France}
\author{I. Bouchoule}
\affiliation{Laboratoire Charles Fabry, Institut d'Optique, CNRS, Universit\'e Paris
Sud 11, 2 Avenue Augustin Fresnel, F-91127 Palaiseau Cedex, France}

\begin{abstract}
We investigate the cooling produced by a  loss process non selective in energy on a one-dimensional (1D) 
Bose gas with repulsive contact interactions in the quasi-condensate regime. 
By performing nonlinear classical field calculations for a homogeneous
system, we show that the gas reaches a non-thermal state where
different modes have acquired different temperatures. After losses have been turned off,
this state is
robust with respect to the nonlinear dynamics, described by the Gross-Pitaevskii
equation. We argue that the integrability of the Gross-Pitaevskii equation
is linked to  the existence of such long-lived non-thermal states, and
illustrate this by showing that such states are not supported within a
non-integrable model of two coupled 1D gases of different masses.  
We go beyond a classical field analysis, taking into account the quantum
noise introduced by the discreteness of losses, and show that the
non-thermal state is still produced and its non-thermal character is even enhanced. Finally, we
extend the discussion to  gases trapped in a harmonic potential and present experimental
observations of a long-lived non-thermal state within a trapped 1D
quasi-condensate following an  atom loss process.
\end{abstract}

\date{\today}
\maketitle

Ultracold temperatures are routinely obtained in dilute atomic gas
experiments using evaporative cooling. Here, an energy-selective loss
process removes the most energetic atoms; provided these atoms have a
high enough energy, rethermalization of the remaining atoms leads to a
lower temperature~\cite{luiten_kinetic_1996}.  Na\"ively, one expects
evaporative cooling to be highly inefficient in (quasi-)
one-dimensional (1D) geometries where the transverse degrees of
freedom are suppressed and the atoms mainly populate the transverse
ground state. Evaporative cooling then only relies on longitudinal
dynamics, and we expect its efficiency to be poor, particularly for
the very shallow longitudinal confinements realized experimentally.
Despite this issue, cooling deep in the 1D regime to temperatures as
low as one tenth of the transverse energy gap has been reached
experimentally
in Bose gas experiments ~\cite{hofferberth_probing_2008,jacqmin_sub-poissonian_2011}. 
This has allowed the realization of 1D
quasi-condensates, where the repulsive interactions between atoms
strongly suppress the density fluctuations and low excitations of the
gas are collective density waves, also called phonons~\cite{kheruntsyan_pair_2003}.  The
nature of the cooling mechanism in such 1D geometries is still not
well understood. However, its investigation is essential in order to
properly characterize both the equilibrium and out-of-equilibirum
properties of these atomic clouds, especially with a view towards
their application in quantum simulation
experiments~\cite{bloch_many-body_2008}.

Recently, Ref.~\cite{grisins_degenerate_2016} theoretically considered a 1D quasi-condensate subject to a 
simple energy-independent loss process 
and showed, within a linearized approach where excitations are treated independently, that cooling was possible. 
The temperature decrease predicted by this theory was observed in an experiment 
probing the low-energy modes of a quasi-condensate undergoing a continuous and homogeneous outcoupling process 
\cite{rauer_cooling_2016}. 
However, studies for homogeneous systems show that
the cooling rate is expected to depend on the mode energy, with 
higher-energy modes cooled at a slower rate than low-energy excitations. 
Thus, as long as the linearized approach is trusted,  losses should produce a 
non-thermal state ({\it i.e.} a 
state that is not described by the Gibbs ensemble).  
Typically, this state is not guaranteed to be long-lived, since coupling 
between modes \textit{a priori} redistributes energy, leading to global thermal equilibrium. 
However, 1D Bose gases
with repulsive contact interactions 
are peculiar since they are described by the Lieb-Liniger Hamiltonian, which 
belongs to the class of 
integrable models. Relaxation of observables towards their values predicted by 
the Gibbs ensemble is not granted in such 
systems~\cite{deutsch_quantum_1991,rigol_thermalization_2008}. 
Consequently, the non-thermal nature of the state 
produced by the loss process could be robust against coupling between modes. 
This might be the origin of the non-thermal nature of the long-lived 1D-quasicondensates 
produced by evaporative cooling and reported 
in~\cite{fang_quench-induced_2014,fang_momentum-space_2016}.
    
In this article, we go beyond the linearized approach and show that
a simple uniform loss process realizes long-lived non-thermal
states of 1D quasi-condensates. We
numerically investigate the simple case of homogeneous gases and
describe the quasi-condensate within a classical field approach,
its dynamics being governed by a nonlinear partial differential equation: the Gross-Pitaevskii equation with an additional term taking losses into account.
We believe the realization of long-lived non-thermal states is related to 
the integrability of the 
system, supported by numerical simulations showing that the system thermalizes towards the 
Gibbs ensemble when integrability is violently broken. 
We then present numerical studies showing that 
long-lived non-thermal states are also produced if one incorporates 
the shot-noise associated with the loss process, 
due to the discreteness
of losses, namely the quantum nature of the atomic field operator.

Finally, we discuss the case of a  gas trapped in a harmonic potential. 
Both the excitation spectrum and the form of the excitations differ from that of an homogeneous system, and
hence one cannot directly extend the results for the homogeneous
system to the trapped case. We nevertheless argue that we still expect a non-thermal state to be produced by the loss process.
We present recent observations of
long-lived out-of-equilibrium states on our experimental atom-chip
setup, that could be related to the conclusions of our theoretical study.

\emph{Linearized approach for homogeneous systems within the classical field approach ---}
We first recall results obtained within the linearized 
approach in the classical field framework.
For this purpose, consider the simple case of a 
 1D Bose gas confined in a box of length $L$ that is initially at 
 thermal equilibrium at  temperature $T_i$ and mean density $\rho_i$.
We use the density/phase representation of the atomic field  
$\psi=\sqrt{\rho}e^{i\theta}$ and denote $\rho_0$ the (time-dependent) mean density.
Density fluctuations $\delta \rho=\rho-\rho_0$ are small in the 
quasi-condensate regime, and phase fluctuations occur on long wavelengths; therefore
as a first approximation one can linearize the equations of motion. 
Expanding $\theta$ and $\delta \rho$ on sinusoidal modes, 
$\theta =\sum_{k>0}\sqrt{2/L} [ \theta_{ck}\cos(kz)+\theta_{sk}\sin(kz)]$ and 
$\delta \rho =\sum_{k>0}\sqrt{2/L} [ \delta\rho_{ck}\cos(kz)+\delta\rho_{sk}\sin(kz)]$,
we find that $\theta_{jk}$ and $\delta\rho_{jk}$ are conjugate variables ({\it i.e.} 
$[\delta\rho_{jk},\theta_{j'k'}]=i \delta_{jj'}\delta_{kk'}$) and that
each mode is governed by its own Hamiltonian 
\begin{equation}
H_{jk}=A_k\delta \rho_{jk}^2+B_k \theta_{jk}^2,
\label{eq.HamBogo}
\end{equation}
where the coefficients $A_k={g/2}+{\hbar^2 k^2}/({8m \rho_0})$ and 
$B_k=\hbar^2k^2\rho_0/(2m)$ depend on $\rho_0$. Here
 $j = c$ or $s$ and $k$ takes discrete values $2n \pi/L$ where $n$ is a positive integer.
Within the classical field approach, the thermal state
of the mode $jk$ corresponds to a Gaussian
distribution of $\theta_{jk}$ and $\rho_{jk}$ satisfying 
the equipartition relation 
$A_k\langle \delta \rho_{jk}^2\rangle =B_k\langle \theta_{jk}^2\rangle =k_B T/2$.

Now consider the uniform loss of atoms at rate $\Gamma$ and 
 its effect on a given mode $jk$.
Losses decrease  $\rho_0 $ at the rate $\Gamma$ and, 
 ignoring at first evolution under the Hamiltonian~(\ref{eq.HamBogo}), 
$\delta\rho_{jk}$ is decreased at the same rate - 
{\it i.e.} $d\delta \rho_{jk}/dt|_{L} = -\Gamma  \delta \rho_{jk}$, where the symbol $L$ 
indicates that we are only considering effect of losses.
Thus the losses decrease the energy in each quadrature, due 
 both to the decrease
of $\delta \rho_{jk}$ and the 
modification of $A_k$ and $B_k$.
If the loss rate is small compared to the mode 
frequency $\omega_k=2\sqrt{A_kB_k}$, one expects 
adiabatic following under the modification of $A_k$ and $B_k$.
In particular,  equipartition 
of energy between the two conjugate variables holds at all times.   
Then, the quantity $\tilde{E}=E/(\hbar\omega_k)$ is 
unaffected by the modification of  $A_k$ and $B_k$ due to the decrease of $\rho_0$, and 
its modification comes solely from the decrease of $\delta\rho_{jk}$ due to losses.
We finally find
\begin{equation}
\frac{d\tilde{E}}{dt}=-\Gamma \tilde{E}.
\label{eq.dEgene}
\end{equation} 
Our assumption of energy equipartition allows us to associate a temperature $k_B T_k = E_k$ 
to the mode, and so Eq.~(\ref{eq.dEgene}) can be rewritten as
\begin{equation}
\frac{T_k(t)}{T_i}= e^{-\Gamma t} \frac{\omega_k(t)}{\omega_k(0)}.
\label{eq.domega}
\end{equation}
Note that the form of Hamiltonian~(\ref{eq.HamBogo}) is not particular to the case 
of a homogeneous gas, provided $\delta \rho_{k,j}$ and $\theta_{k,j}$ are replaced 
by the proper quadratures of the Bogoliubov mode, corresponding to density 
and phase fluctuations, respectively, and $A_k$ and $B_k$ take values which depend on the
Bogoliubov wavefunctions~\cite{wade_manipulation_2016}.
Thus Eq.~(\ref{eq.dEgene}) and (\ref{eq.domega}) are  general, providing the 
adiabatic following condition is fulfilled.
For the particular case of a homogeneous gas, Eq.~(\ref{eq.domega}) gives
\begin{equation}
\frac{dT_k}{dt}=-\Gamma  T_k
\frac{3+\hbar^2k^2/(2\rho_0 m g)}{2+\hbar^2k^2/(2\rho_0 m g)}.
\label{eq.Tcc}
\end{equation}
Losses thus lead to the cooling of each mode, but
at different rates explicitly dependent on $k$.
In the phononic regime $k\ll\sqrt{mgn}/\hbar$, the cooling rate is 
$3\Gamma/2$, compared to $\Gamma$ in the particle 
regime $k\gg\sqrt{mgn}/\hbar$.
Therefore, within the linearized approximation, a uniform loss process produces a non-thermal 
state, where different
modes correspond to different temperatures. Such a state can be viewed as a generalised Gibbs 
ensemble~\cite{langen_experimental_2015}, 
where the different conserved quantities are the energies in each 
linearised mode.

\emph{Nonlinear classical field approach ---}
Beyond the linearized approximation, but still within the classical 
field approach, the system's evolution in the absence of loss is given by  
the Gross Pitaevskii equation for the atomic field $\psi$
\begin{equation}
i\hbar \frac{\partial \psi}{\partial t} = -\frac{\hbar^2}{2m}\frac{\partial^2 \psi}{\partial z}
+g|\psi|^2\psi.
\label{eq.GP}
\end{equation}
This equation contains coupling between the linearized modes studied
above, which acquire a finite 
lifetime~\cite{kulkarni_finite-temperature_2013,mazets_dephasing_2009}.  
 In a generic system, such coupling redistributes the
energy between the modes such that the system reaches the Gibbs
ensemble where all modes share the same temperature.  However, the
Gross-Pitaevskii equation for a 1D homogeneous gas leads to integrable dynamics and
relaxation towards thermal equilibrium is not granted. Consequently, the
out-of-equilibrium state produced by the atom loss process might be
robust against this nonlinear mode coupling.

\begin{figure}[h!]
\centerline{\includegraphics[scale=1, clip, trim = 0.1cm 0.6cm 0 0]{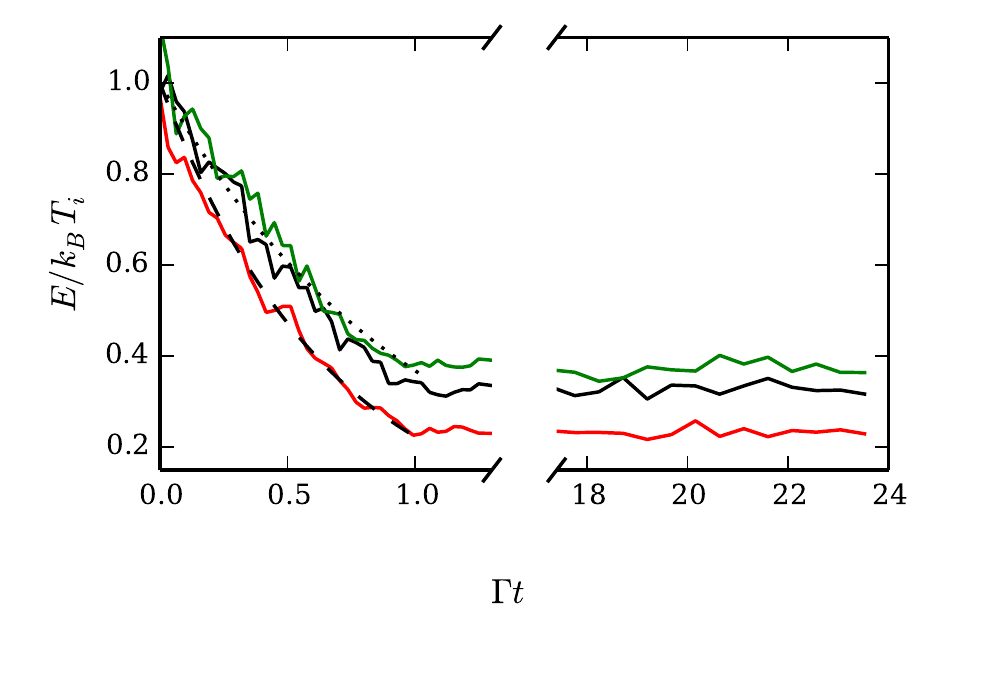}}
\caption{Time evolution of the mode energy during the loss process ($ \Gamma t < 1$), 
and subsequent dynamics after the loss rate is set to zero, for modes 
of wavevectors $k\xi_i=0.5$ (red, lowest curve), 2.5 (black) and 6 (green, highest curve). 
The dashed and dotted lines are the expected behavior 
for phonons ($e^{-3\Gamma t/2}$) and high-energy excitations ($e^{-\Gamma t}$), respectively.
Here  $\chi=T_i/(\rho_i\sqrt{\hbar\rho_i g/m})=0.05$ and  
the loss rate is $\Gamma=2\times 10^{-3} \rho_i g/\hbar $.
}
\label{fig.evolCC}
\end{figure}

To check whether the non-thermal state survives coupling between modes,
we numerically evolved stochastic samples of $\{ \psi(z) \}$   from an
initial thermal state at temperature $T_i$ and density $\rho_i$
according to the dissipative Gross-Pitaevskii equation
(Eq.~(\ref{eq.GP}) with the additional loss term
$\partial\psi/\partial t=-i \Gamma\psi/2$). Each sample ({\it i.e.} each 
single stochastic realization of the initial field $\{ \psi(z) \}$),  was
constructed using the linearized approach above and the associated
thermal Gaussian distribution of the conjugate variables $\theta_{jk}$
and $\delta\rho_{jk}$. Normalizing $\psi$ by $\sqrt{\rho_i}$ and
lengths by $\xi_i=\hbar/\sqrt{mg\rho_i}$, the initial statistical
properties of $\psi$ depend on the single parameter $\chi=T_i/T_{\rm{co}}$
where $T_{\rm{co}}=\hbar\rho_i\sqrt{\rho_i
  g/m}$~\cite{castin_coherence_2000,
  bouchoule_finite-temperature_2016}, while the subsequent time
evolution only depends on $\Gamma/(\rho_i g)$, provided time is
normalized to $\hbar/(\rho_i g)$.  After a certain time $t$, the
quantities $\delta\rho_{jk}$ and $\theta_{jk}$ are extracted, from
which we compute the energy in each mode.  Fig~\ref{fig.evolCC}
shows the time evolution of the mean energy, 
using an ensemble of 10 stochastic  samples, 
in three different Bogoliubov modes of wavevectors $k =
0.5/\xi_i$, $6/\xi_i$, and $2.5/\xi_i$, lying respectively in the
phononic regime, the particle regime, and an intermediate regime.
Here $\chi=0.05$ and the loss rate $\Gamma=2\times 10^{-3}\rho_i g$ is small compared
to the frequencies of the modes analyzed. We verified that
equipartition between the two quadratures is fulfilled within a few
percent during the whole time evolution, confirming that the energy in
each mode can be associated with a temperature.  We find that, for
modes lying in the phononic regime and in the particle regime, the
results are in good agreement with the linearized prediction given by
Eq.~(\ref{eq.Tcc})
and the different modes reach different temperatures.  
This non-thermal situation produced by atom loss is stable over long times; after the loss process has stopped, the
temperature of each mode is stationary
over times as large as $10^3\hbar/\rho_i g$.

Such long-lived non-thermal states are probably only possible due to the 
integrability of the 1D Gross-Pitaevskii equation. 
Nevertheless, the long-lived nature of the state is not obvious, since 
the energies in the linearized modes are not conserved quantities. Lifetimes
of the linearized modes
 are finite~\cite{kulkarni_finite-temperature_2013} and
non-thermal distributions inside 
the phononic regime show good 
thermalization~\cite{grisins_thermalization_2011}. 
 The long lifetime of the non-thermal state generated here is probably 
due to the poor coupling between modes lying in 
the phononic and particle regimes respectively.
The quantum counterpart might be viewed as a form of many-body 
localization in momentum space.

\emph{Effect of integrability on non-thermal state lifetime---} We
investigated the role integrability plays in supporting these
long-lived non-thermal states by considering a closely-related
non-integrable system. Specifically, we coupled a second atomic field
$\varphi$, consisting of particles with mass $m' \neq m$, to the first
field via coupling constant $\tilde{g}$, which is described by the
evolution equations
\begin{equation}
\begin{array}{l}
i\hbar \partial \psi/\partial t = -\frac{\hbar^2}{2m}\frac{\partial^2 \psi}{\partial z}
+(\tilde{g}|\varphi|^2 +  g |\psi|^2)\psi,\\
i\hbar \partial \varphi/\partial t = -\frac{\hbar^2}{2m'}\frac{\partial^2 \varphi}{\partial z}
+(g |\varphi|^2 + \tilde g |\psi|^2)\varphi.\\
\end{array}
\label{eq.GPcoupled}
\end{equation}  
As before, we constructed samples of an initial thermal state by
identifying the two Bogoliubov modes for each wavevector $k$, and
stochastically sampling Gaussian distributions of these modes (for
details see Appendix~\ref{appendix_coupled_BEC}).  We then evolved the
system in presence of losses at the same rate $\Gamma$ for both
species until a substantial fraction of atoms was lost, and
subsequently evolved the system further without the loss term. The
energy in each mode was then extracted via the linearized approach. As
illustrated in Fig.~\ref{fig.evolCC}, when the two fields are coupled
($\tilde g \neq 0$) the modes evolve towards an equipartition of
energy over a long propagation timescale. In contrast, within the
uncoupled system ($\tilde g = 0$) the energies of the modes remain
distinct.

\begin{figure}[h!]
\centerline{\includegraphics[width=8cm,viewport=140 460 430 667,clip]{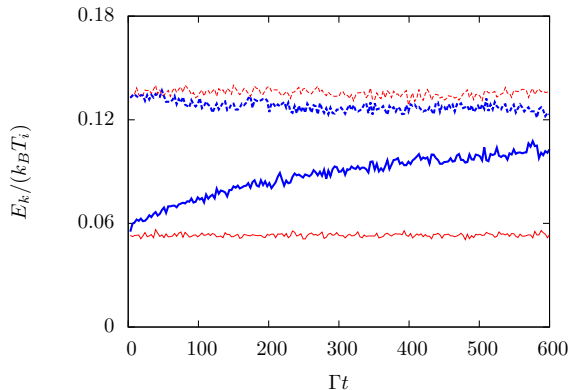}}
\caption{ Time evolution of the mean energy 
  of modes of wave vectors $k=0.3 \sqrt{mg\rho_i}/\hbar$ (solid lines) and $k=6.0\sqrt{mg\rho_i}/\hbar$ 
(dashed lines),
where $\rho_i$ is the initial density of each species, in the
  coupled (fat blue lines, $\tilde{g}=0.4 g$) and uncoupled (thin red lines, $\tilde{g}=0$) cases. 
  The loss process is turned off at time  $t=2/\Gamma$ and
  shown is the subsequent evolution of the isolated system.
  The modes in the two
  uncoupled gases retain their respective energies after dissipation has been turned off
  to form a long-lived non-thermal state as above, while the modes in
  the coupled system relax towards an equipartition of energy. This
  highlights the role of integrability in the establishment of the
  long-lived non-thermal configuration. Here results are obtained by averaging over 10 
  samples, and for each sample we average the mode energy over a $k$ interval of
  0.2$\sqrt{mg\rho_i}/\hbar$. The parameters of the simulation are: $T_i/(\rho_i \sqrt{\hbar g \rho_i/m})=0.04$,
 $m'/m=3$ and $\hbar \Gamma/(\rho_i g)=4\times 10^{-3}$.}
 \label{fig.evolCCcoupled}
\end{figure}

There are many ways to break the integrability of the system. 
In the model of two gases with different masses, the integrability is violently broken 
since a two particle collisional event does not preserve the set of momenta. 
A gentler way to break the integrability would be to consider  two 
gases with atoms of the same mass, but with an interspecies coupling $\tilde{g}$ 
different from the intra-species coupling $g$.
Here, any two-particle collision does preserve the set of momenta.
This system  is nevertheless non-integrable.
However, our simulations of the classical field version of this 
system did not show any relaxation on the time scales shown in Fig~(\ref{fig.evolCC}). 

\emph{Effect of quantum fluctuations associated with the atom-loss process---}
The above treatment does not take into account the quantized nature of the atomic field, {\it i.e.}
the discreteness of the atoms.
In particular, it ignores the shot noise in the loss process,
which introduces additional heating and therefore limits the lowest attainable temperature. 
A description that accounts for the discreteness of the losses
is provided by the stochastic Gross-Pitaevskii equation
\begin{equation}
 i \hbar d \psi=\left ( -\frac{\hbar^2}{2m}\frac{\partial^2}{\partial z^2}\psi
 + g|\psi|^2 \psi  -i\frac{\Gamma}{2} \psi\right ) dt +d \xi,
\label{eq.GPstocastic}
\end{equation}
where $\langle d\xi^*(z)d\xi(z')\rangle = \Gamma dt
\delta(z-z')/2$. This equation can be derived by
converting the master equation for the system density operator to a
partial differential equation for the Wigner quasiprobability
distribution. After the third- and higher-order derivatives 
associated with the nonlinear atomic interaction term are truncated
(an uncontrolled approximation, but one that is typically valid for
weakly-interacting Bose gases, provided the occupation per mode is not
too small over the simulation timescale), evolution of the Wigner
distribution takes the form of a Fokker-Planck equation, which can be
efficiently simulated via Eq.~(\ref{eq.GPstocastic}). 
There exists a formal correspondence between the quantum field
$\hat{\psi}(z)$ and $\psi(z)$: averaging over solutions to
Eq.~(\ref{eq.GPstocastic}) correspond to symmetrically-ordered
expectations (for more details see Appendix~B; an alternative
derivation is provided in \cite{grisins_degenerate_2016}).

As shown in Appendix B, linearizing Eq.~(\ref{eq.GPstocastic}) in
density fluctuations and phase gradient gives an independent evolution
of each mode. Modes with frequencies much larger than the loss rate
remain thermal, however their temperatures depend on the mode energy
and have the following long-time behavior:
$T_{\rm{phonon}}\underset{t\rightarrow \infty}{\simeq}\rho_0(t) g / k_B$
for phononic modes and $T_{\rm{part}}\underset{t\rightarrow
  \infty}{\simeq} \frac{\hbar^2 k^2}{2 m} \frac{1}{k_B \Gamma t}$ for
particle modes. Note that, in contrast to pure classical field predictions,  
the temperature within the particle regime 
depends on $k$. Moreover, the ratio between $T_{\rm{part}}$ and  $T_{\rm{phonon}}$,
\begin{equation}
\frac{T_{\rm{part}}}{T_{\rm{phonon}}}\underset{t\rightarrow \infty}{\simeq}
 \frac{k^2}{2g\rho_i} \frac{e^{\Gamma t}}{\Gamma t},
\label{eq.asympstocastic}
\end{equation}
is much larger
than the one predicted by a pure classical field 
theory. Thus, the effect of the shot noise associated with the discreteness of lost
atoms amplifies the non-thermal nature of the state. 

\begin{figure}
\includegraphics[viewport=145 512 364 661,clip]{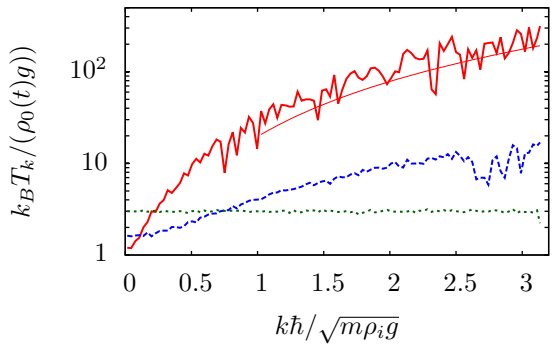}
\caption{Temperature of each mode, obtained from the stochastic Gross-Pitaevskii equation, 
as a function of the wavevector $k$ of the mode,
for different evolution times~: $\Gamma t =0$ (dotted green line), 
$\Gamma t=2.5$ (dashed blue line) and $\Gamma t=5.3$ (solid fat red line). 
The initial temperature is $k_B T_i/(\rho_i g) = 3$. 
As atoms are lost, the gas is driven out of equilibrium and $T_k$ acquire a 
$k$-dependence. 
For phononic particles, we observe that $T_k$ tends towards $g\rho_0(t)$, as expected from the linearized 
approach.
The smooth red solid line is the asymptotic result of Eq.~(\ref{eq.asympstocastic}), valid at long times for  
excitations in the particle regime, computed 
for $\Gamma t =5.3$.
Parameters of the simulation are : $k_B T_i/(\hbar\rho_i\sqrt{g\rho_i/m})=3\times 10^{-3}$, 
$\hbar\Gamma/(g\rho_i)=2\times 10^{-3}$ and $mg/(\hbar^2 \rho_i)=10^{-6}$.}
\label{fig.Tvsk}
\end{figure}

In order to test whether the above predictions including quantum noise
are robust beyond the linearized approach, we numerically simulated
the evolution given by Eq.~(\ref{eq.GPstocastic}).  The initial
thermal state, deep in the quasi-condensate regime, was sampled
stochastically by using the linearized approach and taking into
account quantum fluctuations (which is equivalent to sampling the
Wigner function for a thermal state \cite{Sinatra:2002,
  Olsen:2009}). These samples were then evolved according to
Eq.~(\ref{eq.GPstocastic}), and the energy $E_k$ of each Bogoliubov
mode computed at each time (with averages over trajectories yielding
$\langle E_k\rangle$ and the corresponding temperature
$T_k=\hbar\omega_k/\{ k_B\ln[ (E_k+\hbar\omega_k/2) / (E_k-\hbar\omega_k/2)]\}$.
Fig.~\ref{fig.Tvsk} shows $T_k$ as a function of $k$ at three distinct
times, and reveals that a non-thermal state is realized with a $k$-dependent temperature.  
At small $k$ we find, in agreement
with the linearized approach, that the temperature converges towards
$\rho_0(t)g$ at long times. At long times and large $k$, prediction fron Eq~(\ref{eq.asympstocastic})
are recovered.

\emph{Long-lived non-thermal states in harmonically-confined 1D gases
  ---} The generation of a state which is out-of-equilibrium raises
concerns about experiments probing one-dimensional Bose gases, 
where this non-selective cooling scheme is expected
to occur.  In standard experiments, atoms are confined in a harmonic
potential, which complicates the picture.
To zeroth order in fluctuations, the density profile of the 
gas is given by the Thomas-Fermi inverted parabolic shape~\cite{pethick2008}  \footnote{We assume here the trapping frequency is much smaller than
$g\rho_p$, where $\rho_p$ is the central atomic density.}. 
At finite temperature, excitation modes above this Thomas-Fermi profile get populated. 
If the loss rate is sufficiently small, one expects that each mode 
adiabatically follows the changes of the Thomas-Fermi shape, such that 
each mode can be treated independently and, 
within the pure classical field approximation,
Eq~(\ref{eq.domega}) is still valid, where $k$ is now a positive integer that indexes the mode.
The frequency of phononic modes, {\it i.e.} modes of energy much
smaller than the chemical potential $\mu$, 
are well approximated by
$\omega_k= \nu \sqrt{k(k+1)/2}$, where  $\nu$ is the harmonic trapping frequency~\cite{ho_quasi_1999}.    
Thus, for modes which stay within the phonon regime during the entire loss 
process, Eq.~(\ref{eq.domega}) predicts 
that their temperature decreases as $e^{-\Gamma t}$.

The description of higher-energy modes, called particle modes, is not simple since they explore 
regions where the Thomas-Fermi density vanishes and the quasi-condensate 
approximation fails. It is reasonable however to believe that 
the energy spectrum at energies much larger than $\mu$ is close to the energy spectrum 
of free particles, so that frequencies of these modes are equally spaced, separated by 
$\nu$. 
Since the chemical potential decreases during the loss process, many excitations initially in the
phononic regime are transferred to the particle regime.
Let us consider such an excitation. Its frequency goes from $\omega_k \simeq k\nu/\sqrt{2}$
before the loss process \footnote{We assume here $k\gg 1$.}, to about $k\nu$ at the end of the loss process
when it lies in the particle regime. The ratio
$\omega_k(t_f)/\omega_k(0)$ is thus larger than one.
According to the classical field prediction of Eq.~(\ref{eq.domega}), one therefore expects 
these excitations to attain a higher temperature than those 
lower excitations staying within the phonon regime. 

The effect of shot noise on the loss process is not easy to treat for a trapped gas.
However, we expect that, as in the case of a homogeneous gas, 
the quantum noise will amplify the non-thermal behavior of the system,
so  the temperature differences between modes could be even larger.

\emph{Experimental observation of a long-lived non-thermal state ---}
Observing the non-thermal nature of the gas experimentally requires the ability to address modes
of different energies independently.  This is {\it a priori} not an
easy task for gases confined in a box since all modes overlap
spatially. However, since the atomic clouds in typical experiments are
confined longitudinally in a slowly varying harmonic potential, there 
is some spatial separation of modes of different energy. At
very low temperatures, thermal excitations of
energy larger than $\rho_p g$ give the density profile `wings' that 
extend beyond the Thomas-Fermi inverted parabola of peak density $\rho_{p}$.  
In contrast, low-energy excitations lying in the phononic regime do not
extend beyond the Thomas-Fermi profile, but are responsible for
long wavelength density fluctuations in the central region of the cloud.  
The density profile of the gas is thus most sensitive to high-energy excitations. 
Low-energy excitations, on the other hand, can be probed by
investigating, within the Thomas-Fermi profile, atom-number
fluctuations $\langle \delta N^2 \rangle$, in pixels of length $\Delta$ much
larger than the healing length $\xi_0$~\cite{esteve_observations_2006}.  

\begin{figure*}
\includegraphics[scale = 0.9, clip, trim = 0.4cm 0 0 0]{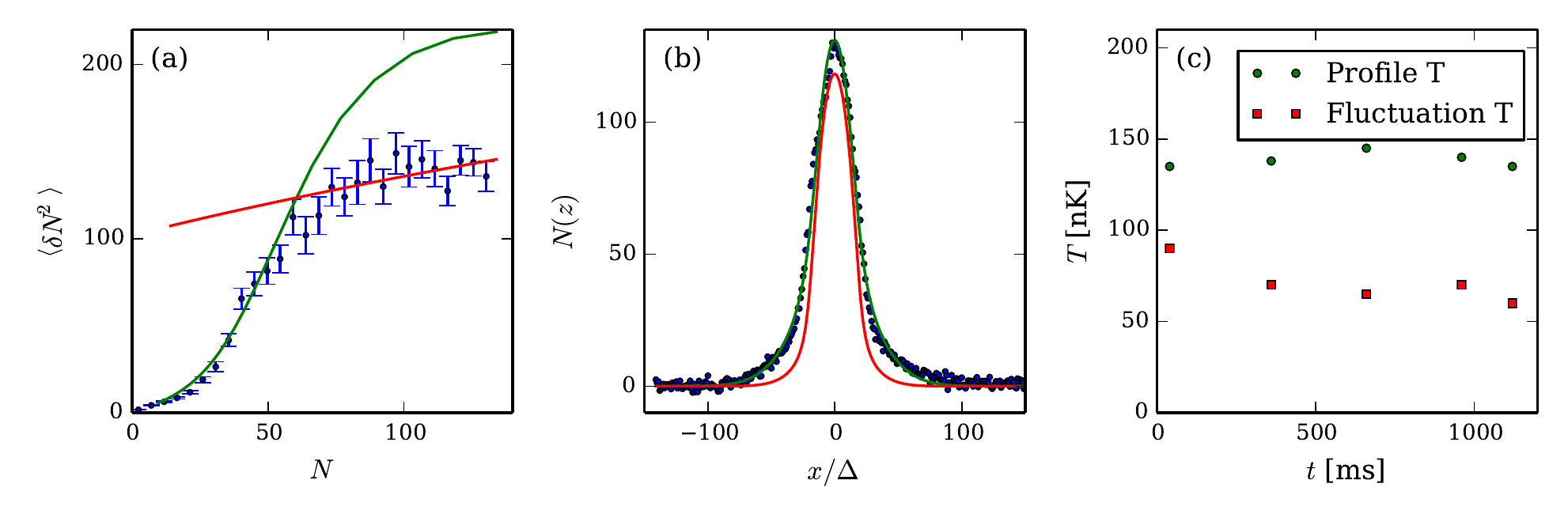}
\caption{Results of the two thermometries we performed on the
  experiment. (a) Atom number fluctuations. Fitting the central region
  of the cloud, {\it i.e.}~the high atom-number part, (red
  line) gives a first temperature $T_{fl} = 80$ nK. However, the
  expected profile from this temperature (also the red curve on (b))
  lies well below the experimental data. A second temperature $T_{pr}
  = 140$ nK is found by fitting the profile with the equation of state
  (solid green line). By plotting the expected atom number
  fluctuations from $T_{pr}$ on (a), the prediction is in good
  agreement with the small atom number region. That is, the center of
  the cloud appears to be at a different temperature than the
  edges. The close to 50\% discrepancy is well beyond the uncertainty
  we have on the temperature measurements, which is around 10 \%.(c)
  Data showing that this temperature difference is stationary over
  time: we observed a long-lived non-thermal state.}
\label{fig.all_exp}
\end{figure*}

Experimentally, we prepare
clouds of $^{87}$Rb atoms by radio-frequency evaporation in our
atom-chip experiment, as described in ~\cite{armijo_mapping_2011}, and we record a set of density
profiles taken under the same experimental conditions. The longitudinal
trapping frequency is 6.2 Hz, while the transverse confinement is 1.9 kHz. 
Atoms are polarized in the $|F=2,m=2\rangle$ hyperfine ground state,
where the interactions are characterized by the $s$-wave scattering length
$a=5.2~nm$. Since the local density
approximation is well fulfilled longitudinally, the equilibrium profile
can be computed using the equation of state for longitudinally
homogeneous gases, $\rho(\mu,T)$, where $\mu$ is the chemical
potential.  Using the well-established modified Yang-Yang equation of
state~\cite{van_amerongen_yang-yang_2008,armijo_mapping_2011}, where
the effective 1D coupling constant is $g=2\hbar\omega_\perp a$, the
experimental density profile is fitted for a temperature
$T_{pr}=140~$nK (see Fig.~\ref{fig.all_exp}).  We also extract
atom-number fluctuations $\langle \delta N^2 \rangle$ in each
pixel from the same dataset, giving an independent temperature measurement.  
Since $\Delta$ is both much smaller than the cloud size and
much larger than the healing length, the physics of homogeneous gases
is locally probed and thermodynamics predicts $\langle \delta N^2 \rangle=k_B T \Delta
\partial \rho /\partial \mu$~\cite{armijo_mapping_2011}.  In
Fig.~\ref{fig.all_exp}, we plot $\langle \delta N^2 \rangle$ versus the mean atom number
in the pixel. Fitting the large atom-number region, corresponding to
pixels lying inside the Thomas-Fermi profile, with the fluctuation-dissipation
relation and the quasi-condensate equation of state, we extract a
temperature $T_\text{fl}=80$~nK (as summarized in Fig.~\ref{fig.all_exp}). The difference between
$T_\text{pr}$ and $T_\text{fl}$ is a signature that the cloud is out-of-equilibrium. 
We also confirmed that, after the radio-frequency
loss mechanism has been removed, this situation is stable over the
cloud lifetime of about one second (Fig.~\ref{fig.all_exp}).  Since the profile is more
sensitive to high-energy excitations while the density fluctuations
are more sensitive to low energy excitations, the fact that
$T_\text{pr}>T_\text{fl}$ could be related to the above quantitative study of homogeneous gases and 
the qualitative arguments given for the trapped system.
In the experiments presented in \cite{rauer_cooling_2016}, only low-energy excitations were probed 
and consequently this non-thermal character was not revealed.


To conclude, we theoretically investigated the long-lived non-thermal
state produced by the non-selective removal of atoms in order to cool
a uniform one-dimensional Bose gas. This dissipation drives the system
out-of-equilibrium, with different excitation modes losing energies at
different rates. This out-of-equilibrium character is robust against
coupling between modes introduced in the Gross-Pitaevskii equation,
and is related to the integrable nature of the considered system.  
We performed simulations of a two-species Bose
mixture with different masses, a non-integrable system, and confirmed a slow relaxation
towards an equipartition of energy between excitations. Truncated
Wigner simulations that go beyond the pure classical-field description
and include the shot-noise associated to the loss process
due to the quantized nature of the atomic field
further confirmed the non-thermal nature of the state produced by
dissipation.  Finally, we discussed the relevance of our findings for
experimental realizations of  1D Bose gases trapped in a harmonic potential. 
From a 
theoretical point of view, in the linearized classical field approach, a small
temperature difference between modes of different energies is indeed
expected, and this effect could be amplified by the presence of
quantum noise.  In our quasi-condensate experiments, we indeed have
signatures of a non-thermal character  since  different
thermometries that probe different parts of the excitation spectrum give substantially
different temperatures. However, a more careful and quantitative
description in the trap, perhaps via finite-temperature classical
field simulations \cite{Blakie:2008}, is still required in order to
draw firm conclusions on the relation between these experimental
long-lived non-thermal states and our theoretical findings.

\emph{Acknowledgements ---} We acknowledge fruitful discussions with
M.~J.~Davis, K.~V.~Kheruntsyan, and M. Olshanii.
This work has been supported by Cnano IdF.  S.S.S. acknowledges
support from the Australian Research Council (ARC) Centre of
Excellence for Engineered Quantum Systems (Grant No. CE110001013) and
the ARC Discovery Project Grant No. DP160103311.


\begin{thebibliography}{0}%
\makeatletter
\providecommand \@ifxundefined [1]{%
 \@ifx{#1\undefined}
}%
\providecommand \@ifnum [1]{%
 \ifnum #1\expandafter \@firstoftwo
 \else \expandafter \@secondoftwo
 \fi
}%
\providecommand \@ifx [1]{%
 \ifx #1\expandafter \@firstoftwo
 \else \expandafter \@secondoftwo
 \fi
}%
\providecommand \natexlab [1]{#1}%
\providecommand \enquote  [1]{``#1''}%
\providecommand \bibnamefont  [1]{#1}%
\providecommand \bibfnamefont [1]{#1}%
\providecommand \citenamefont [1]{#1}%
\providecommand \href@noop [0]{\@secondoftwo}%
\providecommand \href [0]{\begingroup \@sanitize@url \@href}%
\providecommand \@href[1]{\@@startlink{#1}\@@href}%
\providecommand \@@href[1]{\endgroup#1\@@endlink}%
\providecommand \@sanitize@url [0]{\catcode `\\12\catcode `\$12\catcode
  `\&12\catcode `\#12\catcode `\^12\catcode `\_12\catcode `\%12\relax}%
\providecommand \@@startlink[1]{}%
\providecommand \@@endlink[0]{}%
\providecommand \url  [0]{\begingroup\@sanitize@url \@url }%
\providecommand \@url [1]{\endgroup\@href {#1}{\urlprefix }}%
\providecommand \urlprefix  [0]{URL }%
\providecommand \Eprint [0]{\href }%
\providecommand \doibase [0]{http://dx.doi.org/}%
\providecommand \selectlanguage [0]{\@gobble}%
\providecommand \bibinfo  [0]{\@secondoftwo}%
\providecommand \bibfield  [0]{\@secondoftwo}%
\providecommand \translation [1]{[#1]}%
\providecommand \BibitemOpen [0]{}%
\providecommand \bibitemStop [0]{}%
\providecommand \bibitemNoStop [0]{.\EOS\space}%
\providecommand \EOS [0]{\spacefactor3000\relax}%
\providecommand \BibitemShut  [1]{\csname bibitem#1\endcsname}%
\let\auto@bib@innerbib\@empty
\end{thebibliography}%


\begin{thebibliography}{33}%
\makeatletter
\providecommand \@ifxundefined [1]{%
 \@ifx{#1\undefined}
}%
\providecommand \@ifnum [1]{%
 \ifnum #1\expandafter \@firstoftwo
 \else \expandafter \@secondoftwo
 \fi
}%
\providecommand \@ifx [1]{%
 \ifx #1\expandafter \@firstoftwo
 \else \expandafter \@secondoftwo
 \fi
}%
\providecommand \natexlab [1]{#1}%
\providecommand \enquote  [1]{``#1''}%
\providecommand \bibnamefont  [1]{#1}%
\providecommand \bibfnamefont [1]{#1}%
\providecommand \citenamefont [1]{#1}%
\providecommand \href@noop [0]{\@secondoftwo}%
\providecommand \href [0]{\begingroup \@sanitize@url \@href}%
\providecommand \@href[1]{\@@startlink{#1}\@@href}%
\providecommand \@@href[1]{\endgroup#1\@@endlink}%
\providecommand \@sanitize@url [0]{\catcode `\\12\catcode `\$12\catcode
  `\&12\catcode `\#12\catcode `\^12\catcode `\_12\catcode `\%12\relax}%
\providecommand \@@startlink[1]{}%
\providecommand \@@endlink[0]{}%
\providecommand \url  [0]{\begingroup\@sanitize@url \@url }%
\providecommand \@url [1]{\endgroup\@href {#1}{\urlprefix }}%
\providecommand \urlprefix  [0]{URL }%
\providecommand \Eprint [0]{\href }%
\providecommand \doibase [0]{http://dx.doi.org/}%
\providecommand \selectlanguage [0]{\@gobble}%
\providecommand \bibinfo  [0]{\@secondoftwo}%
\providecommand \bibfield  [0]{\@secondoftwo}%
\providecommand \translation [1]{[#1]}%
\providecommand \BibitemOpen [0]{}%
\providecommand \bibitemStop [0]{}%
\providecommand \bibitemNoStop [0]{.\EOS\space}%
\providecommand \EOS [0]{\spacefactor3000\relax}%
\providecommand \BibitemShut  [1]{\csname bibitem#1\endcsname}%
\let\auto@bib@innerbib\@empty
\bibitem [{\citenamefont {Luiten}\ \emph {et~al.}(1996)\citenamefont {Luiten},
  \citenamefont {Reynolds},\ and\ \citenamefont
  {Walraven}}]{luiten_kinetic_1996}%
  \BibitemOpen
  \bibfield  {author} {\bibinfo {author} {\bibfnamefont {O.~J.}\ \bibnamefont
  {Luiten}}, \bibinfo {author} {\bibfnamefont {M.~W.}\ \bibnamefont
  {Reynolds}}, \ and\ \bibinfo {author} {\bibfnamefont {J.~T.~M.}\ \bibnamefont
  {Walraven}},\ }\href {\doibase 10.1103/PhysRevA.53.381} {\bibfield  {journal}
  {\bibinfo  {journal} {Physical Review A}\ }\textbf {\bibinfo {volume} {53}},\
  \bibinfo {pages} {381} (\bibinfo {year} {1996})}\BibitemShut {NoStop}%
\bibitem [{\citenamefont {Hofferberth}\ \emph {et~al.}(2008)\citenamefont
  {Hofferberth}, \citenamefont {Lesanovsky}, \citenamefont {Schumm},
  \citenamefont {Imambekov}, \citenamefont {Gritsev}, \citenamefont {Demler},\
  and\ \citenamefont {Schmiedmayer}}]{hofferberth_probing_2008}%
  \BibitemOpen
  \bibfield  {author} {\bibinfo {author} {\bibfnamefont {S.}~\bibnamefont
  {Hofferberth}}, \bibinfo {author} {\bibfnamefont {I.}~\bibnamefont
  {Lesanovsky}}, \bibinfo {author} {\bibfnamefont {T.}~\bibnamefont {Schumm}},
  \bibinfo {author} {\bibfnamefont {A.}~\bibnamefont {Imambekov}}, \bibinfo
  {author} {\bibfnamefont {V.}~\bibnamefont {Gritsev}}, \bibinfo {author}
  {\bibfnamefont {E.}~\bibnamefont {Demler}}, \ and\ \bibinfo {author}
  {\bibfnamefont {J.}~\bibnamefont {Schmiedmayer}},\ }\href {\doibase
  10.1038/nphys941} {\bibfield  {journal} {\bibinfo  {journal} {Nat Phys}\
  }\textbf {\bibinfo {volume} {4}},\ \bibinfo {pages} {489} (\bibinfo {year}
  {2008})}\BibitemShut {NoStop}%
\bibitem [{\citenamefont {Jacqmin}\ \emph {et~al.}(2011)\citenamefont
  {Jacqmin}, \citenamefont {Armijo}, \citenamefont {Berrada}, \citenamefont
  {Kheruntsyan},\ and\ \citenamefont
  {Bouchoule}}]{jacqmin_sub-poissonian_2011}%
  \BibitemOpen
  \bibfield  {author} {\bibinfo {author} {\bibfnamefont {T.}~\bibnamefont
  {Jacqmin}}, \bibinfo {author} {\bibfnamefont {J.}~\bibnamefont {Armijo}},
  \bibinfo {author} {\bibfnamefont {T.}~\bibnamefont {Berrada}}, \bibinfo
  {author} {\bibfnamefont {K.~V.}\ \bibnamefont {Kheruntsyan}}, \ and\ \bibinfo
  {author} {\bibfnamefont {I.}~\bibnamefont {Bouchoule}},\ }\href {\doibase
  10.1103/PhysRevLett.106.230405} {\bibfield  {journal} {\bibinfo  {journal}
  {Phys. Rev. Lett.}\ }\textbf {\bibinfo {volume} {106}},\ \bibinfo {pages}
  {230405} (\bibinfo {year} {2011})}\BibitemShut {NoStop}%
\bibitem [{\citenamefont {Kheruntsyan}\ \emph {et~al.}(2003)\citenamefont
  {Kheruntsyan}, \citenamefont {Gangardt}, \citenamefont {Drummond},\ and\
  \citenamefont {Shlyapnikov}}]{kheruntsyan_pair_2003}%
  \BibitemOpen
  \bibfield  {author} {\bibinfo {author} {\bibfnamefont {K.}~\bibnamefont
  {Kheruntsyan}}, \bibinfo {author} {\bibfnamefont {D.}~\bibnamefont
  {Gangardt}}, \bibinfo {author} {\bibfnamefont {P.}~\bibnamefont {Drummond}},
  \ and\ \bibinfo {author} {\bibfnamefont {G.}~\bibnamefont {Shlyapnikov}},\
  }\href@noop {} {\bibfield  {journal} {\bibinfo  {journal} {Phys. {Rev}.
  {Lett}.}\ }\textbf {\bibinfo {volume} {91}},\ \bibinfo {pages} {040403}
  (\bibinfo {year} {2003})}\BibitemShut {NoStop}%
\bibitem [{\citenamefont {Bloch}\ \emph {et~al.}(2006)\citenamefont {Bloch},
  \citenamefont {Dalibard},\ and\ \citenamefont
  {Zwerger}}]{bloch_many-body_2008}%
  \BibitemOpen
  \bibfield  {author} {\bibinfo {author} {\bibfnamefont {I.}~\bibnamefont
  {Bloch}}, \bibinfo {author} {\bibfnamefont {J.}~\bibnamefont {Dalibard}}, \
  and\ \bibinfo {author} {\bibfnamefont {W.}~\bibnamefont {Zwerger}},\
  }\href@noop {} {\bibfield  {journal} {\bibinfo  {journal} {{Rev}.
  {Mod}.{Phys}.}\ }\textbf {\bibinfo {volume} {80}},\ \bibinfo {pages} {885}
  (\bibinfo {year} {2006})}\BibitemShut {NoStop}%
\bibitem [{\citenamefont {Grišins}\ \emph {et~al.}(2016)\citenamefont
  {Grišins}, \citenamefont {Rauer}, \citenamefont {Langen}, \citenamefont
  {Schmiedmayer},\ and\ \citenamefont {Mazets}}]{grisins_degenerate_2016}%
  \BibitemOpen
  \bibfield  {author} {\bibinfo {author} {\bibfnamefont {P.}~\bibnamefont
  {Grišins}}, \bibinfo {author} {\bibfnamefont {B.}~\bibnamefont {Rauer}},
  \bibinfo {author} {\bibfnamefont {T.}~\bibnamefont {Langen}}, \bibinfo
  {author} {\bibfnamefont {J.}~\bibnamefont {Schmiedmayer}}, \ and\ \bibinfo
  {author} {\bibfnamefont {I.~E.}\ \bibnamefont {Mazets}},\ }\href {\doibase
  10.1103/PhysRevA.93.033634} {\bibfield  {journal} {\bibinfo  {journal} {Phys.
  Rev. A}\ }\textbf {\bibinfo {volume} {93}},\ \bibinfo {pages} {033634}
  (\bibinfo {year} {2016})}\BibitemShut {NoStop}%
\bibitem [{\citenamefont {Rauer}\ \emph {et~al.}(2016)\citenamefont {Rauer},
  \citenamefont {Gri{\v s}ins}, \citenamefont {Mazets}, \citenamefont
  {Schweigler}, \citenamefont {Rohringer}, \citenamefont {Geiger},
  \citenamefont {Langen},\ and\ \citenamefont
  {Schmiedmayer}}]{rauer_cooling_2016}%
  \BibitemOpen
  \bibfield  {author} {\bibinfo {author} {\bibfnamefont {B.}~\bibnamefont
  {Rauer}}, \bibinfo {author} {\bibfnamefont {P.}~\bibnamefont {Gri{\v s}ins}},
  \bibinfo {author} {\bibfnamefont {I.}~\bibnamefont {Mazets}}, \bibinfo
  {author} {\bibfnamefont {T.}~\bibnamefont {Schweigler}}, \bibinfo {author}
  {\bibfnamefont {W.}~\bibnamefont {Rohringer}}, \bibinfo {author}
  {\bibfnamefont {R.}~\bibnamefont {Geiger}}, \bibinfo {author} {\bibfnamefont
  {T.}~\bibnamefont {Langen}}, \ and\ \bibinfo {author} {\bibfnamefont
  {J.}~\bibnamefont {Schmiedmayer}},\ }\href {\doibase
  10.1103/PhysRevLett.116.030402} {\bibfield  {journal} {\bibinfo  {journal}
  {Phys. Rev. Lett.}\ }\textbf {\bibinfo {volume} {116}},\ \bibinfo {pages}
  {030402} (\bibinfo {year} {2016})}\BibitemShut {NoStop}%
\bibitem [{\citenamefont {Deutsch}(1991)}]{deutsch_quantum_1991}%
  \BibitemOpen
  \bibfield  {author} {\bibinfo {author} {\bibfnamefont {J.~M.}\ \bibnamefont
  {Deutsch}},\ }\href {\doibase 10.1103/PhysRevA.43.2046} {\bibfield  {journal}
  {\bibinfo  {journal} {Phys. Rev. A}\ }\textbf {\bibinfo {volume} {43}},\
  \bibinfo {pages} {2046} (\bibinfo {year} {1991})}\BibitemShut {NoStop}%
\bibitem [{\citenamefont {Rigol}\ \emph {et~al.}(2008)\citenamefont {Rigol},
  \citenamefont {Dunjko},\ and\ \citenamefont
  {Olshanii}}]{rigol_thermalization_2008}%
  \BibitemOpen
  \bibfield  {author} {\bibinfo {author} {\bibfnamefont {M.}~\bibnamefont
  {Rigol}}, \bibinfo {author} {\bibfnamefont {V.}~\bibnamefont {Dunjko}}, \
  and\ \bibinfo {author} {\bibfnamefont {M.}~\bibnamefont {Olshanii}},\ }\href
  {\doibase 10.1038/nature06838} {\bibfield  {journal} {\bibinfo  {journal}
  {Nature}\ }\textbf {\bibinfo {volume} {452}},\ \bibinfo {pages} {854}
  (\bibinfo {year} {2008})}\BibitemShut {NoStop}%
\bibitem [{\citenamefont {Fang}\ \emph {et~al.}(2014)\citenamefont {Fang},
  \citenamefont {Carleo}, \citenamefont {Johnson},\ and\ \citenamefont
  {Bouchoule}}]{fang_quench-induced_2014}%
  \BibitemOpen
  \bibfield  {author} {\bibinfo {author} {\bibfnamefont {B.}~\bibnamefont
  {Fang}}, \bibinfo {author} {\bibfnamefont {G.}~\bibnamefont {Carleo}},
  \bibinfo {author} {\bibfnamefont {A.}~\bibnamefont {Johnson}}, \ and\
  \bibinfo {author} {\bibfnamefont {I.}~\bibnamefont {Bouchoule}},\ }\href
  {\doibase 10.1103/PhysRevLett.113.035301} {\bibfield  {journal} {\bibinfo
  {journal} {Phys. Rev. Lett.}\ }\textbf {\bibinfo {volume} {113}},\ \bibinfo
  {pages} {035301} (\bibinfo {year} {2014})}\BibitemShut {NoStop}%
\bibitem [{\citenamefont {Fang}\ \emph {et~al.}(2016)\citenamefont {Fang},
  \citenamefont {Johnson}, \citenamefont {Roscilde},\ and\ \citenamefont
  {Bouchoule}}]{fang_momentum-space_2016}%
  \BibitemOpen
  \bibfield  {author} {\bibinfo {author} {\bibfnamefont {B.}~\bibnamefont
  {Fang}}, \bibinfo {author} {\bibfnamefont {A.}~\bibnamefont {Johnson}},
  \bibinfo {author} {\bibfnamefont {T.}~\bibnamefont {Roscilde}}, \ and\
  \bibinfo {author} {\bibfnamefont {I.}~\bibnamefont {Bouchoule}},\ }\href
  {\doibase 10.1103/PhysRevLett.116.050402} {\bibfield  {journal} {\bibinfo
  {journal} {Phys. Rev. Lett.}\ }\textbf {\bibinfo {volume} {116}},\ \bibinfo
  {pages} {050402} (\bibinfo {year} {2016})}\BibitemShut {NoStop}%
\bibitem [{\citenamefont {Wade}\ \emph {et~al.}(2016)\citenamefont {Wade},
  \citenamefont {Sherson},\ and\ \citenamefont
  {Mølmer}}]{wade_manipulation_2016}%
  \BibitemOpen
  \bibfield  {author} {\bibinfo {author} {\bibfnamefont {A.~C.~J.}\
  \bibnamefont {Wade}}, \bibinfo {author} {\bibfnamefont {J.~F.}\ \bibnamefont
  {Sherson}}, \ and\ \bibinfo {author} {\bibfnamefont {K.}~\bibnamefont
  {Mølmer}},\ }\href {\doibase 10.1103/PhysRevA.93.023610} {\bibfield
  {journal} {\bibinfo  {journal} {Phys. Rev. A}\ }\textbf {\bibinfo {volume}
  {93}},\ \bibinfo {pages} {023610} (\bibinfo {year} {2016})}\BibitemShut
  {NoStop}%
\bibitem [{\citenamefont {Langen}\ \emph {et~al.}(2015)\citenamefont {Langen},
  \citenamefont {Erne}, \citenamefont {Geiger}, \citenamefont {Rauer},
  \citenamefont {Schweigler}, \citenamefont {Kuhnert}, \citenamefont
  {Rohringer}, \citenamefont {Mazets}, \citenamefont {Gasenzer},\ and\
  \citenamefont {Schmiedmayer}}]{langen_experimental_2015}%
  \BibitemOpen
  \bibfield  {author} {\bibinfo {author} {\bibfnamefont {T.}~\bibnamefont
  {Langen}}, \bibinfo {author} {\bibfnamefont {S.}~\bibnamefont {Erne}},
  \bibinfo {author} {\bibfnamefont {R.}~\bibnamefont {Geiger}}, \bibinfo
  {author} {\bibfnamefont {B.}~\bibnamefont {Rauer}}, \bibinfo {author}
  {\bibfnamefont {T.}~\bibnamefont {Schweigler}}, \bibinfo {author}
  {\bibfnamefont {M.}~\bibnamefont {Kuhnert}}, \bibinfo {author} {\bibfnamefont
  {W.}~\bibnamefont {Rohringer}}, \bibinfo {author} {\bibfnamefont {I.~E.}\
  \bibnamefont {Mazets}}, \bibinfo {author} {\bibfnamefont {T.}~\bibnamefont
  {Gasenzer}}, \ and\ \bibinfo {author} {\bibfnamefont {J.}~\bibnamefont
  {Schmiedmayer}},\ }\href {\doibase 10.1126/science.1257026} {\bibfield
  {journal} {\bibinfo  {journal} {Science}\ }\textbf {\bibinfo {volume}
  {348}},\ \bibinfo {pages} {207} (\bibinfo {year} {2015})}\BibitemShut
  {NoStop}%
\bibitem [{\citenamefont {Kulkarni}\ and\ \citenamefont
  {Lamacraft}(2013)}]{kulkarni_finite-temperature_2013}%
  \BibitemOpen
  \bibfield  {author} {\bibinfo {author} {\bibfnamefont {M.}~\bibnamefont
  {Kulkarni}}\ and\ \bibinfo {author} {\bibfnamefont {A.}~\bibnamefont
  {Lamacraft}},\ }\href {\doibase 10.1103/PhysRevA.88.021603} {\bibfield
  {journal} {\bibinfo  {journal} {Phys. Rev. A}\ }\textbf {\bibinfo {volume}
  {88}},\ \bibinfo {pages} {021603} (\bibinfo {year} {2013})}\BibitemShut
  {NoStop}%
\bibitem [{\citenamefont {Mazets}\ and\ \citenamefont
  {Schmiedmayer}(2009)}]{mazets_dephasing_2009}%
  \BibitemOpen
  \bibfield  {author} {\bibinfo {author} {\bibfnamefont {I.~E.}\ \bibnamefont
  {Mazets}}\ and\ \bibinfo {author} {\bibfnamefont {J.}~\bibnamefont
  {Schmiedmayer}},\ }\href {\doibase 10.1140/epjb/e2008-00421-5} {\bibfield
  {journal} {\bibinfo  {journal} {Eur. Phys. J. B}\ }\textbf {\bibinfo {volume}
  {68}},\ \bibinfo {pages} {335} (\bibinfo {year} {2009})}\BibitemShut
  {NoStop}%
\bibitem [{\citenamefont {Castin}\ \emph {et~al.}(2000)\citenamefont {Castin},
  \citenamefont {Dum}, \citenamefont {Mandonnet}, \citenamefont {Minguzzi},\
  and\ \citenamefont {Carusotto}}]{castin_coherence_2000}%
  \BibitemOpen
  \bibfield  {author} {\bibinfo {author} {\bibfnamefont {Y.}~\bibnamefont
  {Castin}}, \bibinfo {author} {\bibfnamefont {R.}~\bibnamefont {Dum}},
  \bibinfo {author} {\bibfnamefont {E.}~\bibnamefont {Mandonnet}}, \bibinfo
  {author} {\bibfnamefont {A.}~\bibnamefont {Minguzzi}}, \ and\ \bibinfo
  {author} {\bibfnamefont {I.}~\bibnamefont {Carusotto}},\ }\href
  {http://www.tandfonline.com/doi/abs/10.1080/09500340008232189} {\bibfield
  {journal} {\bibinfo  {journal} {Journal of Modern Optics}\ }\textbf {\bibinfo
  {volume} {47}},\ \bibinfo {pages} {2671} (\bibinfo {year}
  {2000})}\BibitemShut {NoStop}%
\bibitem [{\citenamefont {Bouchoule}\ \emph {et~al.}(2016)\citenamefont
  {Bouchoule}, \citenamefont {Szigeti}, \citenamefont {Davis},\ and\
  \citenamefont {Kheruntsyan}}]{bouchoule_finite-temperature_2016}%
  \BibitemOpen
  \bibfield  {author} {\bibinfo {author} {\bibfnamefont {I.}~\bibnamefont
  {Bouchoule}}, \bibinfo {author} {\bibfnamefont {S.~S.}\ \bibnamefont
  {Szigeti}}, \bibinfo {author} {\bibfnamefont {M.~J.}\ \bibnamefont {Davis}},
  \ and\ \bibinfo {author} {\bibfnamefont {K.~V.}\ \bibnamefont
  {Kheruntsyan}},\ }\href {\doibase 10.1103/PhysRevA.94.051602} {\bibfield
  {journal} {\bibinfo  {journal} {Phys. Rev. A}\ }\textbf {\bibinfo {volume}
  {94}},\ \bibinfo {pages} {051602} (\bibinfo {year} {2016})}\BibitemShut
  {NoStop}%
\bibitem [{\citenamefont {Grisins}\ and\ \citenamefont
  {Mazets}(2011)}]{grisins_thermalization_2011}%
  \BibitemOpen
  \bibfield  {author} {\bibinfo {author} {\bibfnamefont {P.}~\bibnamefont
  {Grisins}}\ and\ \bibinfo {author} {\bibfnamefont {I.~E.}\ \bibnamefont
  {Mazets}},\ }\href {\doibase 10.1103/PhysRevA.84.053635} {\bibfield
  {journal} {\bibinfo  {journal} {Phys. Rev. A}\ }\textbf {\bibinfo {volume}
  {84}},\ \bibinfo {pages} {053635} (\bibinfo {year} {2011})}\BibitemShut
  {NoStop}%
\bibitem [{\citenamefont {Sinatra}\ \emph {et~al.}(2002)\citenamefont
  {Sinatra}, \citenamefont {Lobo},\ and\ \citenamefont
  {Castin}}]{Sinatra:2002}%
  \BibitemOpen
  \bibfield  {author} {\bibinfo {author} {\bibfnamefont {A.}~\bibnamefont
  {Sinatra}}, \bibinfo {author} {\bibfnamefont {C.}~\bibnamefont {Lobo}}, \
  and\ \bibinfo {author} {\bibfnamefont {Y.}~\bibnamefont {Castin}},\ }\href
  {http://stacks.iop.org/0953-4075/35/i=17/a=301} {\bibfield  {journal}
  {\bibinfo  {journal} {Journal of Physics B: Atomic, Molecular and Optical
  Physics}\ }\textbf {\bibinfo {volume} {35}},\ \bibinfo {pages} {3599}
  (\bibinfo {year} {2002})}\BibitemShut {NoStop}%
\bibitem [{\citenamefont {Olsen}\ and\ \citenamefont
  {Bradley}(2009)}]{Olsen:2009}%
  \BibitemOpen
  \bibfield  {author} {\bibinfo {author} {\bibfnamefont {M.}~\bibnamefont
  {Olsen}}\ and\ \bibinfo {author} {\bibfnamefont {A.}~\bibnamefont
  {Bradley}},\ }\href {\doibase 10.1016/j.optcom.2009.06.033} {\bibfield
  {journal} {\bibinfo  {journal} {Optics Communications}\ }\textbf {\bibinfo
  {volume} {282}},\ \bibinfo {pages} {3924 } (\bibinfo {year}
  {2009})}\BibitemShut {NoStop}%
\bibitem [{\citenamefont {Pethick}\ and\ \citenamefont
  {Smith}(2008)}]{pethick2008}%
  \BibitemOpen
  \bibfield  {author} {\bibinfo {author} {\bibfnamefont {C.~J.}\ \bibnamefont
  {Pethick}}\ and\ \bibinfo {author} {\bibfnamefont {H.}~\bibnamefont
  {Smith}},\ }\href {\doibase 10.1017/CBO9780511802850} {\emph {\bibinfo
  {title} {Bose–Einstein Condensation in Dilute Gases:}}},\ \bibinfo
  {edition} {2nd}\ ed.\ (\bibinfo  {publisher} {Cambridge University Press},\
  \bibinfo {address} {Cambridge},\ \bibinfo {year} {2008})\BibitemShut
  {NoStop}%
\bibitem [{Note1()}]{Note1}%
  \BibitemOpen
  \bibinfo {note} {We assume here the trapping frequency is much smaller than
  $g\rho _p$, where $\rho _p$ is the central atomic density.}\BibitemShut
  {Stop}%
\bibitem [{\citenamefont {Ho}\ and\ \citenamefont {Ma}(1999)}]{ho_quasi_1999}%
  \BibitemOpen
  \bibfield  {author} {\bibinfo {author} {\bibfnamefont {T.-L.}\ \bibnamefont
  {Ho}}\ and\ \bibinfo {author} {\bibfnamefont {M.}~\bibnamefont {Ma}},\ }\href
  {\doibase 10.1023/A:1021894713105} {\bibfield  {journal} {\bibinfo  {journal}
  {Journal of Low Temperature Physics}\ }\textbf {\bibinfo {volume} {115}},\
  \bibinfo {pages} {61} (\bibinfo {year} {1999})}\BibitemShut {NoStop}%
\bibitem [{Note2()}]{Note2}%
  \BibitemOpen
  \bibinfo {note} {We assume here $k\gg 1$.}\BibitemShut {Stop}%
\bibitem [{\citenamefont {Esteve}\ \emph {et~al.}(2006)\citenamefont {Esteve},
  \citenamefont {Trebbia}, \citenamefont {Schumm}, \citenamefont {Aspect},
  \citenamefont {Westbrook},\ and\ \citenamefont
  {Bouchoule}}]{esteve_observations_2006}%
  \BibitemOpen
  \bibfield  {author} {\bibinfo {author} {\bibfnamefont {J.}~\bibnamefont
  {Esteve}}, \bibinfo {author} {\bibfnamefont {J.-B.}\ \bibnamefont {Trebbia}},
  \bibinfo {author} {\bibfnamefont {T.}~\bibnamefont {Schumm}}, \bibinfo
  {author} {\bibfnamefont {A.}~\bibnamefont {Aspect}}, \bibinfo {author}
  {\bibfnamefont {C.~I.}\ \bibnamefont {Westbrook}}, \ and\ \bibinfo {author}
  {\bibfnamefont {I.}~\bibnamefont {Bouchoule}},\ }\href@noop {} {\bibfield
  {journal} {\bibinfo  {journal} {Phys. Rev. Lett.}\ }\textbf {\bibinfo
  {volume} {96}},\ \bibinfo {pages} {130403} (\bibinfo {year}
  {2006})}\BibitemShut {NoStop}%
\bibitem [{\citenamefont {Armijo}\ \emph {et~al.}(2011)\citenamefont {Armijo},
  \citenamefont {Jacqmin}, \citenamefont {Kheruntsyan},\ and\ \citenamefont
  {Bouchoule}}]{armijo_mapping_2011}%
  \BibitemOpen
  \bibfield  {author} {\bibinfo {author} {\bibfnamefont {J.}~\bibnamefont
  {Armijo}}, \bibinfo {author} {\bibfnamefont {T.}~\bibnamefont {Jacqmin}},
  \bibinfo {author} {\bibfnamefont {K.}~\bibnamefont {Kheruntsyan}}, \ and\
  \bibinfo {author} {\bibfnamefont {I.}~\bibnamefont {Bouchoule}},\ }\href
  {\doibase 10.1103/PhysRevA.83.021605} {\bibfield  {journal} {\bibinfo
  {journal} {Phys. Rev. A}\ }\textbf {\bibinfo {volume} {83}},\ \bibinfo
  {pages} {021605} (\bibinfo {year} {2011})}\BibitemShut {NoStop}%
\bibitem [{\citenamefont {van Amerongen}\ \emph {et~al.}(2008)\citenamefont
  {van Amerongen}, \citenamefont {van Es}, \citenamefont {Wicke}, \citenamefont
  {Kheruntsyan},\ and\ \citenamefont {van
  Druten}}]{van_amerongen_yang-yang_2008}%
  \BibitemOpen
  \bibfield  {author} {\bibinfo {author} {\bibfnamefont {A.~H.}\ \bibnamefont
  {van Amerongen}}, \bibinfo {author} {\bibfnamefont {J.~J.~P.}\ \bibnamefont
  {van Es}}, \bibinfo {author} {\bibfnamefont {P.}~\bibnamefont {Wicke}},
  \bibinfo {author} {\bibfnamefont {K.~V.}\ \bibnamefont {Kheruntsyan}}, \ and\
  \bibinfo {author} {\bibfnamefont {N.~J.}\ \bibnamefont {van Druten}},\ }\href
  {\doibase 10.1103/PhysRevLett.100.090402} {\bibfield  {journal} {\bibinfo
  {journal} {Phys. Rev. Lett.}\ }\textbf {\bibinfo {volume} {100}},\ \bibinfo
  {pages} {090402} (\bibinfo {year} {2008})}\BibitemShut {NoStop}%
\bibitem [{\citenamefont {Blakie}\ \emph {et~al.}(2008)\citenamefont {Blakie},
  \citenamefont {Bradley}, \citenamefont {Davis}, \citenamefont {Ballagh},\
  and\ \citenamefont {Gardiner}}]{Blakie:2008}%
  \BibitemOpen
  \bibfield  {author} {\bibinfo {author} {\bibfnamefont {P.~B.}\ \bibnamefont
  {Blakie}}, \bibinfo {author} {\bibfnamefont {A.~S.}\ \bibnamefont {Bradley}},
  \bibinfo {author} {\bibfnamefont {M.~J.}\ \bibnamefont {Davis}}, \bibinfo
  {author} {\bibfnamefont {R.~J.}\ \bibnamefont {Ballagh}}, \ and\ \bibinfo
  {author} {\bibfnamefont {C.~W.}\ \bibnamefont {Gardiner}},\ }\bibfield
  {booktitle} {\emph {\bibinfo {booktitle} {Advances in Physics}},\ }\href
  {\doibase 10.1080/00018730802564254} {\bibfield  {journal} {\bibinfo
  {journal} {Advances in Physics}\ }\textbf {\bibinfo {volume} {57}},\ \bibinfo
  {pages} {363} (\bibinfo {year} {2008})}\BibitemShut {NoStop}%
\bibitem [{\citenamefont {Drummond}\ and\ \citenamefont
  {Hardman}(1993)}]{Drummond:1993}%
  \BibitemOpen
  \bibfield  {author} {\bibinfo {author} {\bibfnamefont {P.~D.}\ \bibnamefont
  {Drummond}}\ and\ \bibinfo {author} {\bibfnamefont {A.~D.}\ \bibnamefont
  {Hardman}},\ }\href {http://stacks.iop.org/0295-5075/21/i=3/a=005} {\bibfield
   {journal} {\bibinfo  {journal} {EPL (Europhysics Letters)}\ }\textbf
  {\bibinfo {volume} {21}},\ \bibinfo {pages} {279} (\bibinfo {year}
  {1993})}\BibitemShut {NoStop}%
\bibitem [{\citenamefont {Carter}(1995)}]{Carter:1995}%
  \BibitemOpen
  \bibfield  {author} {\bibinfo {author} {\bibfnamefont {S.~J.}\ \bibnamefont
  {Carter}},\ }\href {\doibase 10.1103/PhysRevA.51.3274} {\bibfield  {journal}
  {\bibinfo  {journal} {Phys. Rev. A}\ }\textbf {\bibinfo {volume} {51}},\
  \bibinfo {pages} {3274} (\bibinfo {year} {1995})}\BibitemShut {NoStop}%
\bibitem [{\citenamefont {Steel}\ \emph {et~al.}(1998)\citenamefont {Steel},
  \citenamefont {Olsen}, \citenamefont {Plimak}, \citenamefont {Drummond},
  \citenamefont {Tan}, \citenamefont {Collett}, \citenamefont {Walls},\ and\
  \citenamefont {Graham}}]{Steel:1998}%
  \BibitemOpen
  \bibfield  {author} {\bibinfo {author} {\bibfnamefont {M.~J.}\ \bibnamefont
  {Steel}}, \bibinfo {author} {\bibfnamefont {M.~K.}\ \bibnamefont {Olsen}},
  \bibinfo {author} {\bibfnamefont {L.~I.}\ \bibnamefont {Plimak}}, \bibinfo
  {author} {\bibfnamefont {P.~D.}\ \bibnamefont {Drummond}}, \bibinfo {author}
  {\bibfnamefont {S.~M.}\ \bibnamefont {Tan}}, \bibinfo {author} {\bibfnamefont
  {M.~J.}\ \bibnamefont {Collett}}, \bibinfo {author} {\bibfnamefont {D.~F.}\
  \bibnamefont {Walls}}, \ and\ \bibinfo {author} {\bibfnamefont
  {R.}~\bibnamefont {Graham}},\ }\href {\doibase 10.1103/PhysRevA.58.4824}
  {\bibfield  {journal} {\bibinfo  {journal} {Phys. Rev. A}\ }\textbf {\bibinfo
  {volume} {58}},\ \bibinfo {pages} {4824} (\bibinfo {year}
  {1998})}\BibitemShut {NoStop}%
\bibitem [{\citenamefont {Polkovnikov}(2010)}]{Polkovnikov:2010}%
  \BibitemOpen
  \bibfield  {author} {\bibinfo {author} {\bibfnamefont {A.}~\bibnamefont
  {Polkovnikov}},\ }\href {\doibase
  http://dx.doi.org/10.1016/j.aop.2010.02.006} {\bibfield  {journal} {\bibinfo
  {journal} {Annals of Physics}\ }\textbf {\bibinfo {volume} {325}},\ \bibinfo
  {pages} {1790} (\bibinfo {year} {2010})}\BibitemShut {NoStop}%
\bibitem [{\citenamefont {Opanchuk}\ and\ \citenamefont
  {Drummond}(2013)}]{Opanchuk:2013}%
  \BibitemOpen
  \bibfield  {author} {\bibinfo {author} {\bibfnamefont {B.}~\bibnamefont
  {Opanchuk}}\ and\ \bibinfo {author} {\bibfnamefont {P.~D.}\ \bibnamefont
  {Drummond}},\ }\href {\doibase 10.1063/1.4801781} {\bibfield  {journal}
  {\bibinfo  {journal} {Journal of Mathematical Physics}\ }\textbf {\bibinfo
  {volume} {54}},\ \bibinfo {eid} {042107} (\bibinfo {year}
  {2013})}\BibitemShut {NoStop}%
\end{thebibliography}
%

\appendix

\section{Breaking integrability : two coupled 1D Bose gases} \label{appendix_coupled_BEC}
An example of a non-integrable system is two quasi-condensates of different masses $m$ and $m'$ 
oupled via an interaction term of coupling constant $\tilde g$. 
Here integrability is broken by two-body collisions involving an atom of each species, which does not preserve the set of two initial momenta.
Within the classical field approximation, this system is described by the Hamiltonian
\begin{equation}
\begin{array}{ll}
H=&\int dz \frac{\hbar^2}{2m}\left |\frac{\partial \psi}{\partial z} \right |^2
+\int dz \frac{\hbar^2}{2m'}\left |\frac{\partial \varphi}{\partial z} \right |^2\\
&+\frac{g}{2}\int dz |\psi(z)|^4
+\frac{g}{2}\int dz |\varphi(z)|^4\\
&+\tilde g \int dz |\psi(z)|^2|\varphi(z)|^2
\end{array}
\label{eq.HcoupledSM}
\end{equation}  
which yields the equations of motion Eqs.~(\ref{eq.GPcoupled}). Within the density/phase representation, we can write $\psi = \sqrt{\rho} e^{i \theta}$ and $\varphi = \sqrt{\tilde{\rho}} e^{i \tilde{\theta}}$.
For sufficiently low temperatures, the repulsive interactions result in very small density fluctuations and
long wavelength phase fluctuations, such that 
one can linearize the equations of motion in $\delta \rho$, $\delta \tilde{\rho}$, $\partial \theta/\partial z $ and 
$\partial \tilde{\theta}/\partial z$, or equivalently retain only second-order terms in the Hamiltonian, which can then be diagonalized using a standard Bogoliubov procedure. We give more details on this approach below.

Since Eqs.~(\ref{eq.GPcoupled}) do not explicitly depend on $z$, different Fourier components evolve independently of each other.
Let us consider the Fourier components of wave vector $k=2n\pi/L$ where $n$ is a positive integer and $L$ is the length of the box that confines the gases. As for the single component case, we introduce Fourier coefficients $\delta \rho_{ck}=\sqrt{2/L}\int dz \delta \rho(z)\cos(kz)$, and 
$\delta {\rho}_{sk}=\sqrt{2/L}\int dz \delta \rho(z)\sin(kz)$, and similarly for 
$\delta \tilde\rho$, $\theta$ and $\tilde\theta_2$. 
Each mode $jk$ evolves independently according to the quadratic Hamiltonian
\begin{equation}
\begin{array}{ll}
H_{jk}=& (\frac{g}{2}+\frac{\hbar^2k^2}{8m\rho_0}) \delta {\rho}_{jk}^2 + \frac{\hbar^2k^2\rho_0}{2m}{\theta}_{jk}^2\\
&+(\frac{g}{2}+\frac{\hbar^2k^2}{8m'\rho_0}) \delta {\tilde\rho}_{jk}^2 + \frac{\hbar^2k^2\rho_0}{2m'}{\tilde\theta}_{jk}^2\\
&+\tilde g \delta {\rho}_{jk}\delta {\tilde\rho}_{jk},
\end{array}
\label{eq.HcoupledlinearSM}
\end{equation}
where $j = c$ or $s$. This gives the following linearized equation of motion   
\begin{equation}
i\hbar\frac{\partial}{\partial t}
\left (
\begin{array}{l}
2\sqrt{\rho_0}{\theta}_{jk}\\
\delta {\rho}_{jk}/\sqrt{\rho_0}\\
2\sqrt{\rho_0}{\tilde\theta}_{jk}\\
\delta {\tilde\rho}_{jk}/\sqrt{\rho_0}\\
\end{array}
\right )
= {\cal L}
\left (
\begin{array}{l}
2\sqrt{\rho_0}{\theta}_{jk}\\
\delta {\rho}_{jk}/\sqrt{\rho_0}\\
2\sqrt{\rho_0}{\tilde \theta}_{jk}\\
\delta {\tilde\rho}_{jk}/\sqrt{\rho_0}\\
\end{array}
\right )
\end{equation}
where 
\begin{equation}
{\cal L}=i
\left (
\begin{array}{cccc}
  0 & 2\rho_0 g +\frac{\hbar^2 k^2}{2m}&0&2\tilde{g}\rho_0\\
-\frac{\hbar^2k^2}{2m}&0&0&0\\
0&2\tilde{g}\rho_0&0&2\rho_0 g +\frac{\hbar^2 k^2}{2m'}\\
0&0&-\frac{\hbar^2k^2}{2m'}&0
\end{array}
\right ).
\label{eqSM.matL}
\end{equation}
Symmetry properties of ${\cal L}$ show that this operator has two eigenvectors
 \begin{equation}
{\cal L}\left ( 
\begin{array}{c}
F_1^+\\
iF_1^-\\
F_2^+\\
iF_2^-\\
\end{array}
\right )
=
\omega_a
\left ( 
\begin{array}{c}
F_1^+\\
iF_1^-\\
F_2^+\\
iF_2^-\\
\end{array}
\right ) 
\end{equation}
and
 \begin{equation}
{\cal L}\left ( 
\begin{array}{c}
G_1^+\\
iG_1^-\\
G_2^+\\
iG_2^-\\
\end{array}
\right )
=
\omega_b
\left ( 
\begin{array}{c}
G_1^+\\
iG_1^-\\
G_2^+\\
iG_2^-\\
\end{array}
\right ) 
\end{equation}
where, $F_1^+,F_1^-,F_2^+,F_2^-,G_1^+,G_1^-,G_2^+,G_2^-$ are real
and satisfy the normalisation condition 
\begin{equation}
\left \{
\begin{array}{l}
G_1^-G_1^++G_2^-G_2^+=1 \\
F_1^-F_1^++F_2^-F_2^+=1.
\end{array}\right .
\end{equation}  The vectors
$(F_1^+,-iF_1^-,F_2^+,-iF_2^-)^T$ and
$(G_1^+,-iG_1^-,G_2^+,-iG_2^-)^T$ are eigenvectors of ${\cal L}$ of
eigenergies $-\omega_a$ and $-\omega_b$ respectively, which complete
the basis.  Expanding the state $(2\sqrt{\rho_0}{\theta}_{jk},\delta
{\rho}_{jk}/\sqrt{\rho_0},2\sqrt{\rho_0}{\tilde \theta}_{jk},\delta
{\tilde \rho}_{jk}/\sqrt{\rho_0})^\top$ 
on these the eigenbasis of 
${\cal L}$ gives
\begin{equation}
\left \{
\begin{array}{lll}
2\sqrt{\rho_0}{\theta}_{jk}&=&-F_1^+ i (a-a^*) - G_1^+ i (b-b^*)\\
\delta {\rho}_{jk}/\sqrt{\rho_0}&=&F_1^- (a+a^*)+G_1^- (b+b^*)\\
2\sqrt{\rho_0}{\tilde \theta}_{jk}&=&-F_2^+ i (a-a^*) - G_2^+ i (b-b^*)\\
\delta {\tilde \rho}_{jk}/\sqrt{\rho_0}&=&F_2^- (a+a^*)+G_2^- (b+b^*)\\\\
\end{array}
\right .
\end{equation}
where $a$ and $b$ are c-numbers satisfying
\begin{subequations}
\begin{align}
	i\hbar\partial a/\partial t	&= \hbar \omega_a a, \\
	i\hbar\partial b/\partial t	&= \hbar \omega_b b.
\end{align}
\end{subequations}
Inserting into Eq.~(\ref{eq.HcoupledlinearSM}), we find that the  Hamiltonian $H_{jk}$ can be written as:
\begin{equation}
H_{jk}=\hbar \omega_a |a|^2 + \hbar \omega_b |b|^2
\end{equation}
Although the above procedure utilizes the classical field approach,
a quantum version yields similar results, with $a$ and $b$ replaced by bosonic operators and
$\hat{H}_{jk}=E_k^0+\hbar \omega_a \hat{a}^\dag \hat{a} + \hbar \omega_b \hat{b}^\dag \hat{b}$,
where $E_k^0$ is the contribution of the modes $a$ and $b$ to the vaccuum energy.

We use the linearization above to sample the initial state according to a thermal distribution.
For this purpose, for each $jk$ Fourier component, we diagonalize ${\cal L}$ and we then sample
the c-number $a$ and $b$ according to the thermal Gaussian law $k_B T =\hbar\omega_a \langle |a|^2 \rangle$
and  $k_B T =\hbar\omega_b \langle |b|^2 \rangle$.
From this, we can compute the fields $\psi$ and $\varphi$, subsequently evolve according to Eqs.~(\ref{eq.GPcoupled}), and extract the energy $H_{jk}$ of each Fourier component at each time point.

\section{Stochastic Gross-Pitaevskii equation}
\subsection{Derivation via truncated Wigner}
Here we present a derivation of Eq.~(\ref{eq.GPstocastic}) from a Wigner distribution formalism and the truncated Wigner approximation. This methodology has had great success in the numerical modelling of weakly-interacting Bose gases in regimes where quantum fluctuations are important \cite{Drummond:1993, Carter:1995, Steel:1998, Polkovnikov:2010}, and furthermore underpins the classical field methodology used for both zero and finite temperature simulations \cite{Blakie:2008}. Since the Bose gas is described by a quantum field, the derivation should strictly rely upon functional calculus (for details see, for example, \cite{Opanchuk:2013}). However, since we are primarily concerned with numerical simulation on discrete grids with a finite number of points, for simplicity of presentation we will discretize the problem. That is, we divide space into cells of length $\delta x$, and discretize the field operator such that $\hat{\psi}_r$ annihilates an atom in the cell $r$, and satisfies $[\hat\psi^\dag_r,\hat\psi_{r'}]=\delta_{r,r'}$. Furthermore, we introduce the 
per-cell interaction energy $\tilde g = g / \delta x$ 
and the operator $ \partial_r^2 \equiv \partial^2 / \partial_r^2$, 
which must be interpreted as $\partial_r^2  \left . \{f\} \right |_r = (f_{r+1}+f_{r-1}-2 f_r) / \delta x^2$ when applied to a discrete function $f_r$, where integer $r$ indexes the cell. 

A homogeneous 1D Bose gas undergoing a non-selective loss process can be described by the master equation
\begin{equation} \label{master_eq}
\frac{\partial \rho}{\partial t}  = -\frac{i}{\hbar}[\hat{H},\rho] + \Gamma \sum_r \mathcal{D}[\hat\psi_r] \rho,
\end{equation}
where $\rho$ is the system density operator, $\mathcal{D}[\hat{L}] \rho \equiv \hat{L} \rho \hat{L}^\dag - \tfrac{1}{2} \hat{L}^\dag \hat{L} \rho - \tfrac{1}{2} \rho \hat{L}^\dag \hat{L}$, and $\hat{H}$ is the Lieb-Liniger Hamiltonian
\begin{equation}
\hat{H}=\sum_r \left ( -\hat\psi_r^\dag \frac{\hbar^2}{2m} \partial_r^2 \hat\psi_r + 
\frac{\tilde g}{2} \hat \psi_r^\dag \hat \psi_r^\dag \psi_r \psi_r \right ).
\end{equation}
The system density operator can be equivalently described by the Wigner quasiprobability distribution, $W$, 
which is a real function of a complex field ${\psi(z)}$ :
\begin{equation}
W( \left\{\psi_r, \psi_r^* \right\})=\int \prod_r d^2\lambda_r \frac{e^{-(\lambda_r \psi^*_r+\lambda^*_r\psi_r)}}{\pi^2}\chi(\left\{\lambda_r,\lambda_r^*\right\}),
\end{equation} 
where $\chi(\left\{\lambda_r,\lambda_r^*\right\})$ is the characteristic function
\begin{equation}
	\chi(\left\{\lambda_r,\lambda_r^*\right\}) = \text{Tr}\left \{ \rho \exp\left[\sum_r(\lambda_r \hat\psi_r^\dag - \lambda_r^* \hat\psi_r)\right] \right \}.
\end{equation} 
Averages of functions of $\psi_r,\psi_r^*$ over $W$ 
correspond to expectations of the corresponding symmetrically-ordered operators. 
Using the operator correspondences \cite{Blakie:2008, Opanchuk:2013}
\begin{subequations}
\begin{align}
	\hat{\psi}_r \rho \to \left( \psi_r + \frac{1}{2} \frac{\partial}{\partial \psi_r^*}\right) W( \left\{\psi_r, \psi_r^* \right\}), \\
	\hat{\psi}_r^\dag \rho \to \left( \psi_r^* - \frac{1}{2} \frac{\partial}{\partial \psi_r}\right) W( \left\{\psi_r, \psi_r^* \right\}), \\
	\rho \hat{\psi}_r  \to \left( \psi_r - \frac{1}{2} \frac{\partial}{\partial \psi_r^*}\right) W( \left\{\psi_r, \psi_r^* \right\}), \\
	\rho \hat{\psi}_r^\dag  \to \left( \psi_r^* + \frac{1}{2} \frac{\partial}{\partial \psi_r}\right) W( \left\{\psi_r, \psi_r^* \right\}),
\end{align}
\end{subequations}
we can map the master equation Eq.~(\ref{master_eq}) to the following partial differential equation for the Wigner function:
\begin{equation} \label{Wigner_PDE}
	\frac{\partial W}{\partial t}	= \frac{\partial W}{\partial t} \Big|_\text{Kin} + \frac{\partial W}{\partial t}\Big|_\text{Nonlin} + \frac{\partial W}{\partial t}\Big|_\text{Loss}
\end{equation}
where
\begin{align}
	\frac{\partial W}{\partial t} \Big|_\text{Kin} &= \frac{i \hbar}{2m}\sum_r \left\{ \frac{\partial}{\partial \psi_r} \partial_r^2 \psi_r   - \frac{\partial}{\partial \psi_r^*}  \partial_r^2 \psi_r^*  \right\} W
\end{align}
corresponds to the kinetic energy term, 
\begin{multline}
	\frac{\partial W}{\partial t} \Big|_\text{Nonlin} 	=\frac{i \tilde g}{\hbar} \sum_r \Bigg\{ \frac{1}{4}\left( \frac{\partial^3}{\partial^2 \psi_r \partial \psi_r^*} \psi_r - \frac{\partial^3}{\partial^2 \psi_r^* \partial \psi_r} \psi_r^* \right) \\
										 + \left(\frac{\partial}{\partial \psi_r} \psi_r   - \frac{\partial}{\partial \psi_r^*} \psi_r^* \right)
 (| \psi_r |^2 -1)     \Bigg\} W, \label{Wigner_nonlin}
\end{multline}
corresponds to the nonlinear atom-atom collisional term, and  
\begin{align}
	\frac{\partial W}{\partial t}\Big|_\text{Loss} &= \frac{\Gamma}{2} \sum_{r}  \left \{ 
-\frac{\partial}{\partial \psi_r} \psi_r + \frac{\partial}{\partial \psi^*_r} \psi^*_r +  \frac{\partial^2}{\partial \psi_r \partial \psi_r^*} \right \} W
\end{align}
corresponds to the loss term.  
 This is currently no easier to simulate
than the master equation Eq.~(\ref{master_eq}). However, if we
truncate the third-order derivatives in term Eq.~(\ref{Wigner_nonlin})
that arise due to the nonlinearity, then Eq.~(\ref{Wigner_PDE}) takes
the form of a Fokker-Planck equation with positive definite
diffusion. It can therefore be efficiently simulated via a set of
stochastic differential equations. 
One can replacing $(|\psi|^2 -1) $ in  equation~(\ref{Wigner_nonlin}) by $|\psi|^2$ 
since this corresponds to a simple, irrelevant, energy shift. 
We then find that the differential equations are just the   the stochastic
Gross-Pitaevskii equation, Eq.~(\ref{eq.GPstocastic}). The truncation
of these third-order derivatives is an uncontrolled approximation, but
is typically valid for weakly-interacting Bose gases, provided the
occupation per mode is not too small over the simulation timescale.
Note that the truncated Wigner approximation applied here concerns the treatment of interactions
between atoms in the quasi-condensate. The sole effect of losses is captured in a 
exact way by this procedure at the quantum level.

\subsection{Linearized approach}
In the quasicondensate regime density fluctuations and phase gradients are small.
A linearized approach can therefore be used to
identify independent modes, following the procedure below.
Separating the real and imaginary parts of Eq.~(\ref{eq.GPstocastic}) and linearizing in density fluctuations and 
the phase gradient gives the stochastic equations
\begin{equation}
\left \{
\begin{array}{l}
d \delta \rho_r = -\frac{\hbar^2 \rho_0}{m} \partial_r^2 \theta dt -\Gamma \delta \rho_r  dt +  \sqrt{\rho_0} d\eta_r \\
d\theta_r =-(\tilde g-\frac{\hbar^2}{4m \rho_0}\partial_r^2 )\delta \rho_r dt +\frac{1}{2\sqrt{\rho_0}} d\nu_r 
\end{array}
\right .
\end{equation}
where $d\nu_r$ and $d\eta_r$ are random Gaussian variables with zero mean and variances $\langle d\eta_r d\eta_{r'} \rangle =
\langle d\nu_r d\nu_{r'} \rangle = \delta_{r,r'}\Gamma dt$.
Expanding $\theta_r$ and $\delta \rho_r$ on sinusoidal modes, 
$\theta =\sum_{k>0}\sqrt{2/L} [ \theta_{ck}\cos(kz)+\theta_{sk}\sin(kz)]$ and 
$\delta \rho =\sum_{k>0}\sqrt{2/L} [ \delta\rho_{ck}\cos(kz)+\delta\rho_{sk}\sin(kz)]$
gives
\begin{equation}
\left \{
\begin{array}{l}
d \delta \rho_{jk} = \frac{\hbar^2 k^2 \rho_0}{m}  \theta_{jk} dt -\Gamma \delta \rho_{jk}dt  +  \sqrt{\rho_0} d\eta_{jk},\\
d\theta_{jk} =-(\tilde g-\frac{\hbar^2k^2}{4m \rho_0})\delta \rho_{jk} dt +\frac{1}{2\sqrt{\rho_0}} d\nu_{jk} ,
\end{array}
\right .
\label{eq.stocasticdeltan}
\end{equation}
where $d\nu_{jk}$ and $d\eta_{jk}$ are random Gaussian variables of
vanishing mean and variances $\langle d\eta_{jk}^2 \rangle = \langle
d\nu_{jk}^2 \rangle = \Gamma dt$.  An initial centered
Gaussian Wigner distribution (such as a thermal state) remains
Gaussian under the above linearized stochastic
evolution. Moreover, after averaging over stochastic
  trajectories, it remains centered on $\langle \delta
  \rho_{jk}\rangle=\langle \theta_{jk}\rangle=0$.  Consequently the
Wigner distribution for each mode is entirely determined by
variances and covariances of the variables - explicitly, entirely
determined by the following coupled differential equations:
 \begin{equation}
\left \{
\begin{array}{l}
\frac{d}{dt} \langle \delta \rho_{jk}^2\rangle  = 2\frac{\hbar^2 k^2 \rho_0}{m}  \langle \theta_{jk} \delta \rho_{jk}\rangle -2\Gamma \langle \delta \rho_{jk}^2\rangle  +  \rho_0\Gamma\\
\frac{d}{dt} \langle \theta_{jk}^2 \rangle =-2(\tilde g-\frac{\hbar^2k^2}{4m \rho_0})\langle \theta_{k,j} \delta \rho_{jk}\rangle +\frac{\Gamma}{4\rho_0} \\
\frac{d}{dt} \langle \delta \rho_{jk}\theta_{jk} \rangle=\frac{\hbar^2k^2 \rho_0}{m}\langle \theta_{jk}^2 \rangle
-(\tilde g-\frac{\hbar^2k^2}{4m \rho_0})\langle \theta_{jk}^2\rangle
\end{array}
\right .
\label{eq.vardeltan2}
\end{equation}
The link between these \emph{classical} averages over $\delta \rho_{jk}$ and $\delta \theta_{jk}$ and the expectations over the corresponding \emph{quantum} operators is not immediate. Strictly, averages over various combinations of the $c$-number fields $\psi_r$ and $\psi_r^*$ correspond to expectations of symmetric orderings of the corresponding quantum operators - for example 
$$\langle (\hat{\psi}_{r'}^\dag \hat{\psi}_r + \hat{\psi}_{r} \psi_{r'}^\dag)/2 \rangle = \overline{\psi_{r'}^* \psi_r}.$$  
However, since in the quasi-condensate regime
correlation lengths are much larger than the mean inter-particle 
distance and  density fluctuations are small, one can use a coarse-grained approximation
where the atom number in each cell is large yielding small relative fluctuations. Then 
the atomic density $\langle\hat{\psi}_{r}^\dag \hat{\psi}_r\rangle$ and its higher-order moments are well-approximated simply by 
classical averages over $\psi^*\psi$ and its powers. Put another way, those corrections that arise due to the non-commutativity of the operators are small and can be neglected. A similar argument holds for the phase operator. Consequently, we are justified in interpreting those classical averages within Eqs.~(\ref{eq.vardeltan2}) as quantum expectations.  

Let us focus on the evolution of a given mode of wavevector $k$ and 
assume the loss rate is 
very small compared to the mode frequency $\omega_k=\sqrt{\hbar^2 k^2/ (2m)(\hbar^2 k^2/ (2m)+2\tilde g \rho_0)}$.
Then, the free evolution ensures equipartition of the 
energy between the two conjugate variables $\delta \rho_{jk}$ and $\theta_{jk}$ at any time, 
which corresponds to thermal equilibrium. 
The Wigner function is then solely determined by the mean energy in the mode $E_k$ and one 
finds, from Eq.~(\ref{eq.vardeltan2}), 
\begin{equation}
\frac{d}{dt}\tilde{E}=\Gamma\left (-\tilde E +(\bar A_k^2+1/\bar A_k^2)/4 \right )
\label{eq.dEtildedt}
\end{equation}
where
$\bar A_k=[(\hbar^2 k^2/(2m)+2g\rho_0)/(\hbar^2 k^2/(2m))]^{1/4}$ and $\tilde E = E_k/\omega_k$. 
For phonons, $\bar A_k \approx [4 m g\rho_0/ (\hbar^2 k^2) ]^{1/4}$ and is much larger than 1.
 Using this time-dependant approximation of $A_k$ to solve Eq.~(\ref{eq.dEtildedt}),
we find that $E_k$ asymptotically goes towards $g\rho_0(t)$.
Since $g\rho_0(t)$ is much larger than the ground state energy for phonons,
the  Rayleigh-Jeans limit is attained, and 
this corresponds to a thermal equilibrium at temperature 
\begin{equation}
k_B T_{\rm{phonon}}\underset{t\rightarrow \infty}{\simeq}\rho_0(t) g.
\end{equation}  
In contrast, for modes with $k \gg \sqrt{\rho_0 g}$, an expansion of 
$\bar A_k$ in power of $g\rho_0/ (\hbar^2 k^2 / m)$, one finds
\begin{equation}
\tilde{\epsilon} \simeq \left ( \frac{m g\rho_0}{ \hbar^2 k^2}\right )^2 e^{-\Gamma t} (1-e^{-\Gamma t}) +\tilde{\epsilon}_0 e^{-\Gamma t}
\label{eq.epsilontilde}
\end{equation}  
where $\tilde\epsilon=(\tilde E -1/2)/\omega_k$ 
is the mean quantum occupation number of the mode.
At large times, $\tilde\epsilon$ becomes much smaller than one. This corresponds to a temperature 
$k_B T \simeq -(\hbar^2 k^2/(2m))\ln(\tilde\epsilon)$, much smaller than $\omega_k$.
At large times, we find 
\begin{equation}
k_B T_{\rm{part}}\underset{t\rightarrow \infty}{\simeq} \frac{\hbar^2 k^2}{2m} \frac{1}{\Gamma t}.
\end{equation}
The temperature of those modes depends on $k$ and takes much larger values than $T_{\rm{phonon}}$.

Finally, note that, while in this appendix we start from the truncated Wigner 
stochastic equation to derive
the above linearized approach, an alternative approach is to linearize
the Lieb-Liniger Hamiltonian and then consider, for a given mode, the effect of losses. 
Thus, the validity of linearized approach does not require that the mode occupation number be large. 
It is valid even in the quantum regime, the approximation here being that the gas lies deeply enough 
in the quasi-condensate regime.

\end{document}